# Orthogonally Polarized Kerr Frequency Combs

Changjing Bao[1*], Peicheng Liao[1], Arne Kordts[2], Lin Zhang[3], Andrey Matsko[4], Maxim Karpov[2], Martin H. P. Pfeiffer[2], Guodong Xie[1], Yinwen Cao[1], Yan Yan[1], Ahmed Almaiman[1], Zhe Zhao[1], Amirhossein Mohajerin-Ariaei[1], Ahmad Fallahpour[1], Fatemeh Alishahi[1], Moshe Tur[5], Lute Maleki[4], Tobias J. Kippenberg[2], Alan E. Willner[1]

[1]Department of Electrical Engineering, University of Southern California, Los Angeles, CA 90089, USA.

[2]École Polytechnique Fédérale de Lausanne (EPFL), CH-1015 Lausanne, Switzerland

[3]School of Precision Instrument and Opto-electronics Engineering, Tianjin University, Tianjin 300072, China

[4]OEwaves Inc., 465 N. Halstead Street, Pasadena, California 91107, USA

[5]School of Electrical Engineering, Tel Aviv University, Ramat Aviv 69978, Israel

Corresponding author: changjib@usc.edu

**Abstract**: Kerr optical frequency combs with multi-gigahertz spacing have previously been demonstrated in chip-scale microresonators, with potential applications in coherent communication, spectroscopy, arbitrary waveform generation, and radio frequency photonic oscillators. In general, the harmonics of a frequency comb are identically polarized in a single microresonator. In this work, we report that one comb in one polarization is generated by an orthogonally polarized soliton comb and two low-noise, orthogonally polarized combs interact with each other and exist simultaneously in a single microresonator. The second comb generation is attributed to the strong cross-phase modulation with the orthogonally polarized soliton comb and the high peak power of the intracavity soliton pulse. Experimental results show that a second frequency comb is excited even when a continuous wave light as a "seed"—with power as low as 0.1 mW—is input, while its own power level is below the threshold of comb generation. Moreover, the second comb has a concave envelope, which is different from the $sech^2$ envelope of the soliton comb. This is due to the frequency mismatch between the harmonics and the resonant frequency. We also find that the repetition rates of these two combs coincide, although two orthogonal resonant modes are characterized by different free spectral ranges.
.

Compact microresonator-based frequency combs have attracted much attention, and they have been demonstrated in various applications, including frequency metrology and communications [1-5]. A Kerr comb can be generated when a coherent continuous wave (CW) light is coupled into a high-quality-factor (Q-factor) resonator and its power reaches the threshold of the nonlinear parametric process [6]. Several operating regimes with different noise properties, including primary, chaotic, breather, and stable soliton combs, have been demonstrated under different conditions [7-18]. The stable soliton comb has the characteristics of low noise [19, 20] and wideband coverage [16], and results from the balance between the anomalous dispersion and the Kerr nonlinearity [20].

Recently, there have been reports on the generation of two combs in a single microresonator [21-23]. For example, it was shown that two counter-propagating soliton combs in the same polarization are generated when two pumps with similar power levels are counter-coupled into a single microresonator [21, 22]. However, in these works, there has been little study on the fundamental physical interaction between the two combs in one microresonator. Also, to our knowledge, no work has been reported that one comb can generate another comb and interacts with each other in a single microresonator.

In this work, we demonstrate that one comb in one polarization can help generate an orthogonally polarized comb via cross-phase modulation (XPM) and that two combs in orthogonal polarization states can exist simultaneously in a single resonator. Specifically, the interaction effect between two polarized modes caused by XPM can alleviate the power requirement for one polarized pump light. Also, after the generation of two combs, the existence of the second comb is dependent on the other vertically polarized comb and there is still strong interaction between two orthogonally polarized frequency combs. In the experiment, we pump the microring resonator first with a transverse magnetic (TM)-polarized CW light which leads to the generation of a soliton comb. Then, after the introduction of a much weaker transverse electric (TE)-polarized CW light as a "seed", a TE-polarized frequency comb is excited. Numerical simulations that consider the group–velocity mismatch between two polarization modes also confirm the experimental observation. Thus, not only do we observe that one comb is generated from an orthogonally polarized soliton comb via XPM, but also two combs in different polarizations in a single microresonator are achieved. This approach could be another feasible method of generating an on-chip dual comb.

**Operation principle** Figure 1 shows the schematic of TE- and TM- polarized comb generation in a microresonator. After a TM-polarized pump is coupled into the microresonator to generate a soliton comb, a weak TE-polarized CW interacts with the TM-polarized comb via XPM and the TE-polarized comb is created.

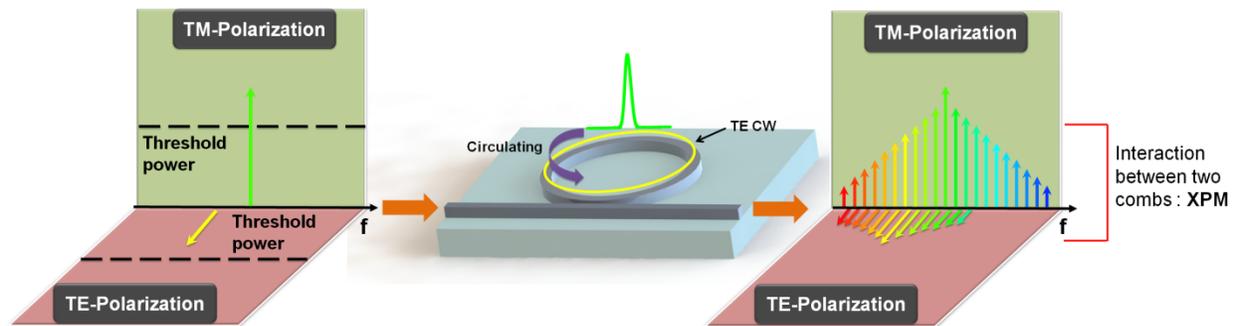

**Figure 1 | TE- and TM- polarized combs in a microresonator.** Schematic of TE-polarized comb generation from a TM-polarized soliton comb. A TE-polarized CW with weak power as a "seed" is cross-phase modulated by a TM-polarized soliton pulse, which results in the generation of a TE-polarized comb.

**Experimental setup and measurement strategy** Figure 2(a) shows our experimental setup. The microresonator in the setup is based on a 1,500-nm height and 900-nm wide silicon nitride waveguide. The radius of the ring resonator is ~119 μm. The TM and TE mode families of the microresonator are characterized by 192.1 GHz and 190.3 GHz free spectral ranges (FSRs), respectively. The microresonator is pumped with the emissions of two external cavity diode lasers (ECDLs) amplified by erbium-doped fiber amplifiers (EDFAs). In the first step, a single-soliton pulse is generated when only a high-power TM-polarized pump laser is coupled into the microresonator. The on-chip pump power is ~600 mW. The pump wavelength is controlled by a function generator, and the comb regime is approached by sweeping the laser from shorter to longer wavelengths across the resonance and stopping at a soliton-existence region [8, 20]. Subsequent laser wavelength backward tuning turns the multi-soliton state into a single soliton state [13] after the observation of a multi-soliton state. In the second step, a TE-polarized CW signal light is also coupled into the cavity. Its on-chip power is ~6.3 mW, a hundred times lower than the power of the first pump. In fact, the power of the TE-polarized signal is lower than the threshold power of the comb generation [6]. Interestingly, we can observe the generation of a TE-polarized comb in the microresonator. The second comb is generated because of the XPM effect between the TE and TM mode families. We also observe that the existence of the second comb relies on the orthogonally polarized soliton comb and the disappearance of the TM-polarized soliton comb also results in the disappearance of the TE-polarized comb.

We use an incoherent optical source to characterize the cavity spectrum. Figure 2(c) shows the measured transmission curves for the fundamental TE and TM modes when a spontaneous-emission source in a single polarization state is coupled into the microresonator. The wavelengths of the TE-polarized signal and TM-polarized pump are 1544.6 nm and 1555.9 nm, respectively, and the corresponding Q factors of the resonant modes are $7.8\times10^5$ and $1.3\times10^6$, respectively. We can see that the TE and TM modes do not overlap in frequency. Figure 2(d) shows the single-soliton comb spectrum when only the TM-polarized pump laser is coupled into the cavity. The comb envelope shows a $sech^2$ shape and the 3 dB comb bandwidth is ~5 THz. The repetition rate is ~191.8 GHz, as seen in Fig. 2(e). The repetition rate of the comb is measured with electro-optical down conversion. The comb is sent into the cascaded optical modulators [24], and the spacing between adjacent electro-optical sidebands is measured.

Figure 2(f) shows the spectrum of both TE- and TM-polarized combs. In addition to the TE-polarized comb, which is generated because of the XPM effect between the TE-polarized CW and the TM-polarized soliton pulse, we can also see a frequency harmonic generated at 1567.4 nm. This line is created because of the degenerate four-wave mixing (FWM) effect between the TE-polarized signal and the TM-polarized pump in the bus waveguide [25]. The wavelength of this line changes when we tune the wavelength of the TE-polarized signal. This line is generated even when the TE-polarized signal is not resonant and the TE-polarized comb is not generated. Figure 2(g) shows the zoom-in spectrum of (f) and we find that the repetition rates of the TE- and TM-polarized combs are identical within the measurement error. Our microresonator has a single-mode "filtering" section [26,27], which is intended to suppress higher-order modes. That is why the TM-polarized frequency comb spectrum in Fig. 2(d) is smooth. The cross-polarization-based XPM interaction occurs only between the fundamental TM and TE modes, and interaction between the fundamental mode and high-order mode families [28,29] is avoided.

To demonstrate the low-phase-noise characteristics of the newly generated TE-polarized comb, we use a narrow-band filter to select the first comb line right next to the TE-polarized signal and the adjacent TM-polarized comb line (see the black-dashed box in Fig. 2[g]). The beat note of these two selected comb lines is measured when they are amplified by an EDFA and sent to a photodetector. A single line peak in the radiofrequency (RF) spectrum of Fig. 2(h) demonstrates that the TE-polarized comb is also in a low-phase-noise state, as is the TM-polarized soliton comb [20]. Furthermore, Fig. 2(i) shows the amplitude noise performances of the two combs, the TM-polarized soliton comb, and the TM-polarized chaotic comb. We can see that the second comb does not introduce a noticeable change to the amplitude noise and that the noise level remains low compared to that of the chaotic comb.

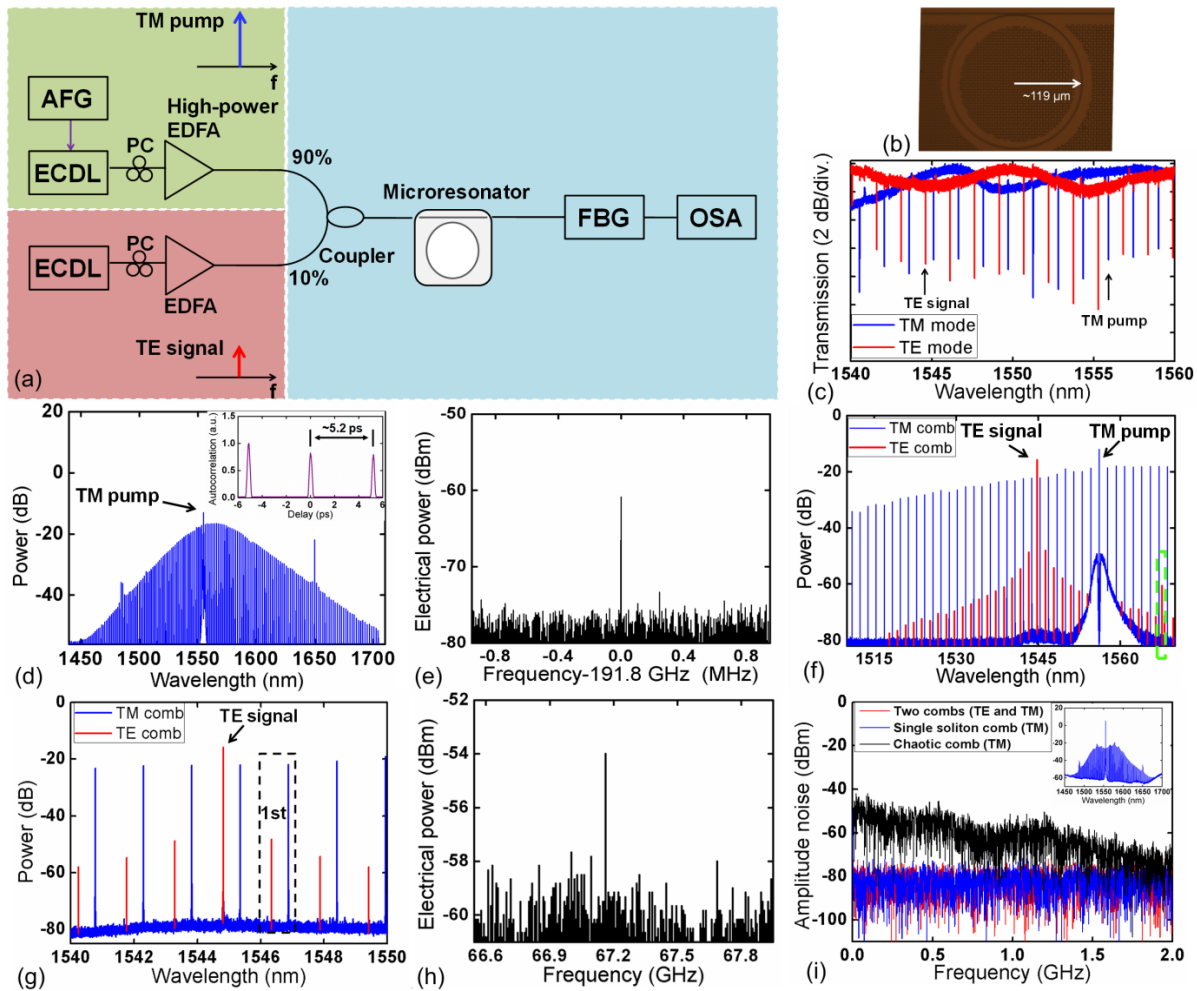

**Figure 2 | Experimental generation of TE- and TM-polarized combs.** (a) Experimental setup of a TE-polarized comb generation when a TE-polarized CW signal with low power is cross-phase modulated by a TM-polarized soliton in the microresonator. The inset shows the intensity autocorrelation of the single soliton. A TM-polarized soliton pulse is generated first by a high-power TM-polarized pump, and a TE-polarized CW signal is then coupled into the cavity as a "seed" to excite the TE-polarized comb. AFG, arbitrary function generator; ECDL, external-cavity diode laser; PC, polarization controller; EDFA, erbium-doped fiber amplifier; FBG, fiber Bragg grating; OSA, optical spectrum analyzer. (b) Optical microscope image of an $Si_3N_4$ microresonator. (c) Measured transmission spectra of the fundamental TE and TM modes when a spontaneous emission source is coupled into the microresonator. (d) Optical spectrum of a single soliton generated by a TM-polarized pump (resolution: 0.1 nm). (e) Repetition rate beat note of the single-soliton comb (resolution bandwidth [RBW] = 10 kHz). (f) Optical spectrum of the TE- and TM-polarized combs (resolution: 0.16 pm). The line at 1567.4 nm in the green-dashed box is generated by the FWM between the TM-polarized pump and the TE-polarized signal on the bus waveguide. (g) Zoom-in spectrum of (f) in a spectral range of 1540 nm to 1550 nm (resolution: 0.16 pm). (h) Heterodyne beat note of the two TE- and TM-polarized comb lines in Fig. 2(g) (RBW = 100 kHz). (i) RF amplitude noises of the two generated combs, the single-soliton comb, and the chaotic comb. The inset shows the spectrum of the chaotic comb.

**Effect of the TE-polarized CW power and the wavelength** Figure 3(a) shows the relationship between the TE-polarized signal power and the power of the generated first sideband right next to the signal. Even at -9.4 dBm (0.1 mW), the TE light generates the second comb, and the corresponding first sideband comb's relative power is -71.3 dB. The ability of the low signal power to excite a frequency comb is a result of the assistance from the TM-polarized soliton pulse and the high peak power of the soliton pulse in the cavity. We also observe the generation of the TE-polarized comb, even when the signal wavelength is moved to other resonant wavelengths. Figures 3(c) and (d) show two example spectra when the signal wavelength is moved to 1549.4 nm and to 1564.9 nm, respectively. Simultaneously, the line directly generated from the FWM between the TE-polarized signal and the TM-polarized pump in the bus waveguide has moved to 1562.5 nm and 1547.0 nm, respectively.

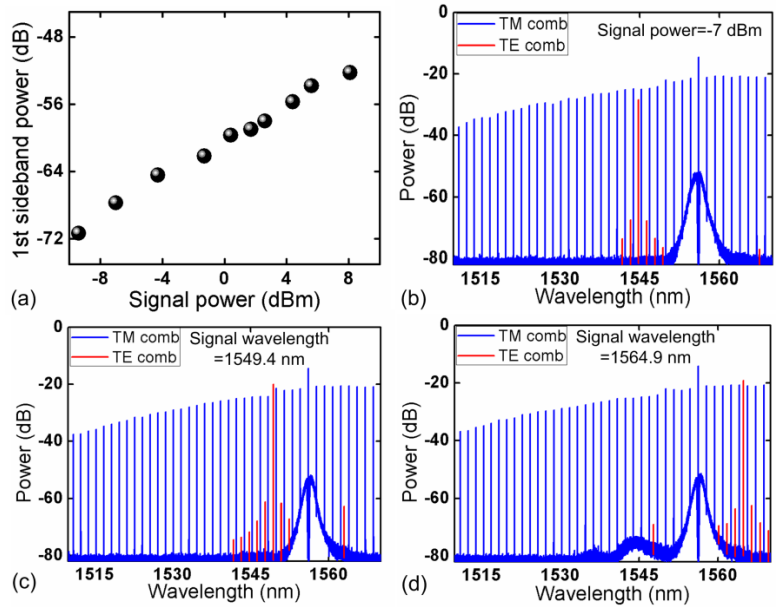

**Figure 3 | TE-polarized comb generation with different input TE-polarized CW powers and wavelengths.** (a) Power of the generated first sideband right next to the TE signal versus the signal power. (b) Sample spectrum of the TE- and TM-polarized combs when the signal power is only -7 dBm. When the signal power is -0.4 dBm and the signal wavelength is moved to 1549.4 nm and 1564.9 nm, the optical spectra of the generated TE- and TM-polarized combs are shown in (c) and (d), respectively (resolution: 0.16 pm).

**XPM between multiple TM-polarized solitons and the TE-polarized CW** The above results experimentally demonstrate that a TE-polarized CW can be modulated by a single-soliton pulse in the microresonator. Moreover, the excitation of multiple solitons at the TM polarization (see Figs. 4[a] and [d]) modifies the TE-polarized comb. This observation indicates that a different number of solitons in the microresonator can assist in generating different TE-polarized comb states. The TM-polarized multi-soliton comb envelope results in the irregularity of the TE-polarized comb envelope.

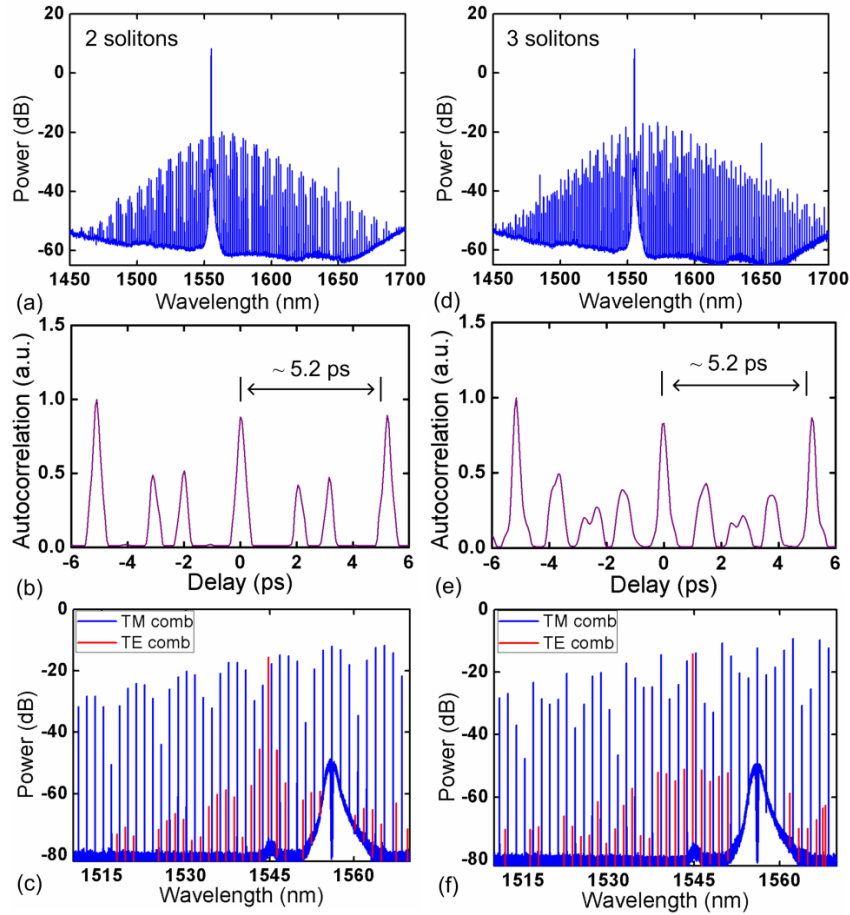

**Figure 4 | TE-polarized comb generation assisted by TM-polarized multisoliton pulses.** The optical spectra of two different multisoliton states are shown in (a) and (d) when only the TM-polarized pump is coupled into the microresonator. The intensity autocorrelations of two multisoliton states are shown in (b) and (e). The corresponding spectra of the TE- and TM-polarized combs are shown in (c) and (f).

**XPM between a single TE-polarized soliton and a TM-polarized CW** A TE-polarized soliton pulse can excite a TM-polarized comb in the same way a TM-polarized pulse excites a TE-polarized comb. Figure 5 shows the spectrum of two combs when the TE-polarized pump wavelength is 1557.1 nm and the TM-polarized signal wavelength is 1545.4 nm. This indicates that a soliton pulse in any polarization state can modulate a CW light in an orthogonal state.

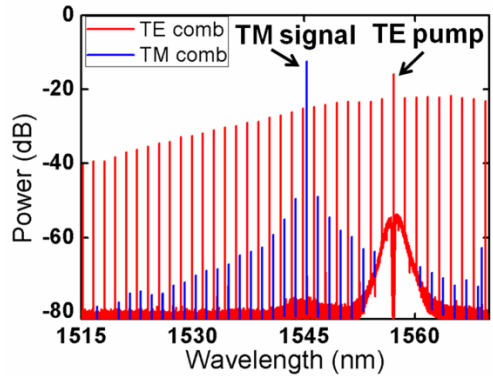

**Figure 5 | TM-polarized comb generation assisted by a TE-polarized soliton comb.** Optical spectrum of two combs, where the TM-polarized comb is generated by XPM between a TE-polarized soliton pulse and a TM-polarized signal. The TE-polarized pump power is 740 mW and the power of the TM-polarized signal is 8.5 mW.

**Numerical Simulations**

We have numerically simulated TE-polarized comb generation via XPM between a TM-polarized soliton pulse and a TE-polarized CW. We consider the XPM effect and the group-velocity mismatch between the TE and TM modes in the modeling. Two coupled Lugiato-Lefever equations [30-33] have been used in the simulation. The second term on the right-hand side of the two equations describes the XPM effect.

$$\left[\tau_0^{TM}\frac{\partial}{\partial t}+\frac{\alpha_i^{TM}}{2}+\frac{\theta^{TM}}{2}-j\delta_0^{TM}+jL\sum_{m=2}^{\infty}\frac{(-j)^m\beta_m^{TM}}{m!}\frac{\partial^m}{\partial\tau^m}\right]E^{TM}(t,\tau)=$$

$$\sqrt{\theta^{TM}}E_{in}^{TM}-j\gamma^{TM}L\cdot E^{TM}(t,\tau)(|E^{TM}(t,\tau)|^2+\frac{2|E^{TE}(t,\tau)|^2}{3}). \quad (1)$$

$$\left[\tau_0^{TE}\frac{\partial}{\partial t}+\frac{\alpha_i^{TE}}{2}+\frac{\theta^{TE}}{2}-j\delta_0^{TE}+jL\sum_{m=2}^{\infty}\frac{(-j)^m\beta_m^{TE}}{m!}\frac{\partial^m}{\partial\tau^m}+L(\frac{1}{v_g^{TM}}-\frac{1}{v_g^{TE}})\frac{\partial}{\partial\tau}\right]E^{TE}(t,\tau)=\sqrt{\theta^{TE}}E_{in}^{TE}-$$

$$j\gamma^{TE}L\cdot E^{TE}(t,\tau)(|E^{TE}(t,\tau)|^2+\frac{2|E^{TM}(t,\tau)|^2}{3}). \quad (2)$$

Here, $\tau_0^{TM}$ and $\tau_0^{TE}$ are the round-trip times for the TM and TE modes, respectively; t and τ are the slow and fast times, respectively; $E^{TM}(t, \tau)$ and $E^{TE}(t, \tau)$ represent the intracavity fields for the two polarization modes, respectively; $E_{in}^{TM}$ and $E_{in}^{TE}$ are the input fields for the two modes, respectively; $\alpha_i^{TM}$ and $\alpha_i^{TE}$ represent the power loss per round trip; $\theta^{TM}$ and $\theta^{TE}$ represent the power coupling coefficient; $\delta_0^{TM}$ and $\delta_0^{TE}$ are the phase detunings of the pumps from the adjacent resonance frequencies, respectively; $\beta_m^{TM}$ and $\beta_m^{TE}$ are the *m*th-order dispersion coefficients; $v_g^{TM}$ and $v_g^{TE}$ are the group velocities; L is cavity length; and $\gamma^{TM}$ and $\gamma^{TE}$ are nonlinearity coefficients. In the simulation, the calculated dispersion coefficients for the TM and TE modes are -129 ps$^2$/km and -105 ps$^2$/km, respectively. The nonlinear coefficients are 0.86/(W·m) and 0.98/(W·m), respectively. The numerical simulations involve two steps to follow the experimental process. In the first step, we use only Eq. 1, in which we ignore the XPM effect to simulate TM-polarized soliton generation. After the soliton becomes stable in the simulation, we then use Eqs. 1 and 2 to simulate both TE-polarized comb generation and TM-polarized comb evolution. We assume that the phase detuning $\delta_0^{TE}$ for the TE mode is zero in the second step.

Figure 6(a) shows the spectral and temporal dynamics of the TM-polarized comb after the TE-polarized CW signal is turned on. The TM-polarized soliton comb evidently exhibits almost no change (see Fig. 6[a], Column 1) because the TE-polarized signal power is much lower than the TM-polarized soliton power, and the XPM effect on the soliton pulse from the TE-polarized signal is consequently weak. On the other hand, we see that the TE-polarized light is modulated, and the corresponding comb is generated. The first column of Fig. 6(a) shows that the TE-polarized comb is generated and it becomes stable within 2 ns. The stable TE-polarized comb spectrum shown in the third column of Fig. 6(a) has a concave envelope, which agrees with the experimental results illustrated in Fig. 2(f). The corresponding temporal waveform in the fourth column of Fig. 6(a) has a dip in the center and does not exhibit a pulse shape. In the pulse-assisted comb generation, several initial comb lines close to the TE-polarized signal are created first, and the comb bandwidth grows until the comb reaches a stable state (see Fig. 6[a], Column 1). We also observe that the temporal evolution trace for the TE-polarized comb has the same slope as that for the TM-polarized comb (see Fig. 6[a], Column 2), which indicates that the repetition rates of the TE- and TM-polarized combs are the same, although the FSR difference for the TE and TM modes is ~1.8 GHz. The TE-polarized comb envelope could be explained as follows. The repetition rate of the TE-polarized comb is significantly different from the FSR of the TE mode. The generation of higher-order harmonics can be suppressed because the further the harmonics are from the TE-polarized CW signal, the larger the frequency mismatch between the stimulated harmonics and the TE mode that accommodates this harmonics.

We also simulate TE-polarized comb generation when the group-velocity mismatch between the TE and TM modes is ignored by assuming $v_g^{TE} = v_g^{TM}$ in Eq. 2. The TM-polarized soliton comb evolution shown in the first column of Fig. 6(b) is similar to that shown in Fig. 6(a). However, the TE-polarized comb exhibits different characteristics (see Fig. 6[b], Line 2). As the third and fourth columns of Fig. 6(b) show, the TE-polarized comb spectrum is broader, and the temporal waveform corresponds to a dark pulse [34]. The pulsewidth is much broader than that of the bright soliton in Fig. 6(b), Column 4, Line 1. Note that this dark pulse is not a standard dark soliton [35], and some oscillation occurs at the pulse edge. A TE-polarized dark pulse can be formed in the anomalous dispersion regime as a result of the strong XPM effect on the signal CW, which counteracts the temporal spreading of the dark pulse induced by the anomalous dispersion [36]. By comparing the TE-polarized comb in Figs. 6(a) and 6(b), we find that even when the group velocities for two modes are close, the small group-velocity mismatch could affect comb generation when the CW is influenced by the soliton pulse.

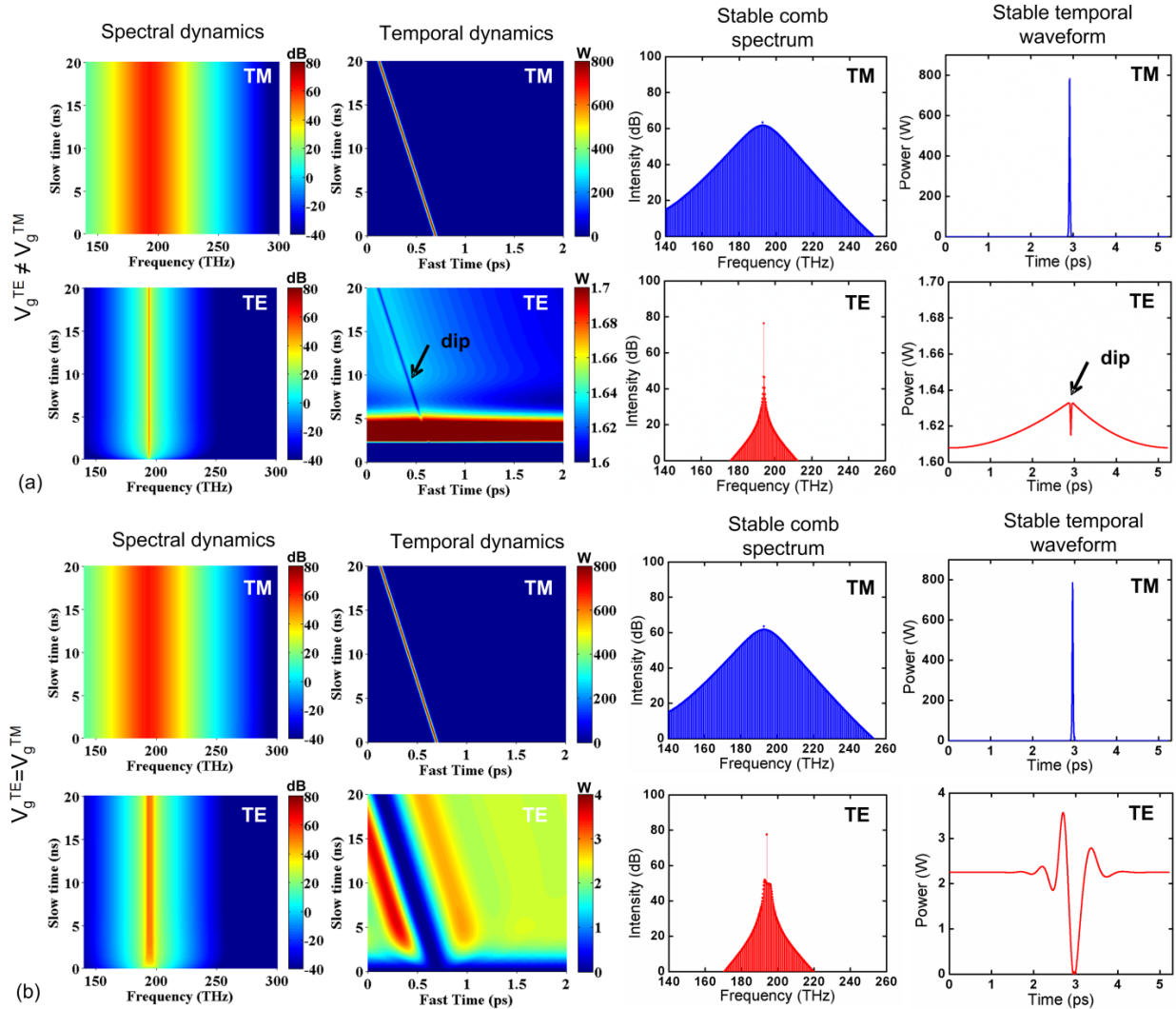

**Figure 6 | Simulated dynamics of TE- and TM-polarized combs.** The simulated spectral and temporal dynamics of the TE- and TM-polarized combs with/without considering the group-velocity mismatch are shown in (a) and (b), respectively. The stable comb spectrum and the corresponding stable temporal waveform are shown in the third and fourth columns, respectively. The temporal waveform of the TE-

polarized comb corresponds to a dark pulse when the group velocities between two polarization modes are equal.

**Discussion**

We have experimentally demonstrated the generation of one polarized frequency comb from an orthogonally polarized soliton comb. This results from the strong XPM effect when a CW signal in one polarization state is modulated by a soliton pulse in an orthogonal polarization state. Because of the high peak power of the soliton pulse in the cavity, even a weak CW signal with low power can excite a comb with the assistance of a soliton. This study may provide another method of generating an on-chip dual comb [3, 18, 21-23, 37-41] in a single microresonator. Several ways could be used to increase the power of the second comb, such as an increase in input signal power, resulting in growth of the power of the second comb. Meanwhile, we should note that the high input power of the signal could disturb the stability of the soliton pulse because of the thermal effect. We could also tune the group velocity difference between two polarization modes by designing the waveguide structure so that the second comb is optimized. In addition, dispersive wave generation [16] could also be used to increase the power and broaden the bandwidth of the second comb.

Furthermore, because there is a nonlinear interaction between two orthogonally polarized frequency combs, the state of one comb could be tuned finely when the other comb is controlled externally. In this work, we use a soliton comb as an example to generate another comb and demonstrate the interaction between two orthogonally polarized Kerr combs. Kerr combs in other states, such as primary combs and breathing soliton combs, may generate different combs in the orthogonally polarization and also interact with the orthogonally polarized combs.

## Methods

**Repetition rate measurement of the TM-polarized soliton comb.** Because the comb spacing of the TM-polarized soliton comb in our microresonator is much greater than the photodetector bandwidth, we use electro-optical down conversion to measure the repetition rate. The generated soliton comb is sent into the cascaded optical modulators composed of an intensity modulator, a phase modulator, and a phase shifter. Each Kerr comb line generates several electro-optic sidebands. The RF signal that drives both modulators is 36 GHz. The bias voltage of the intensity modulator and the RF phase shifter are optimized to flatten the generated electro-optical sidebands. The measured spacing between the second electro-optic sideband from adjacent Kerr comb lines is ~47.8 GHz. Thus, we conclude that the repetition rate of the TM-polarized soliton comb is ~191.8 GHz.

**Acknowledgements**

This work is supported by Air Force Office of Scientific Research (FA9550-15-1-0166), Vannevar Bush Faculty Fellowship program sponsored by the Basic Research Office of the Assistant Secretary of Defense for Research and Engineering and funded by ONR (N00014-16-1-2813), and Defense Security Cooperation Agency (DSCA-4440644029). The authors would like to acknowledge the generous support of Dr. Enrique Parra. Samples were fabricated in the Centre for Micro-Nanotechnology (CMi) at EPFL.

**Author contributions**

The authors declare no competing financial interests. All the authors were involved in the data analysis, and all contributed to writing of the article. C.B. and P.L. performed the experimental measurement. C.B., L.Z. and A.M. performed the numerical simulation. A.K., M.K. and M.H.P.P fabricated the devices. The project was supervised by T.J.K. and A.E.W.